# Miniaturized Photoacoustic Spectroscopy Gas Probe for In-Situ Detection in Oil

Yi Zhong[1], Yuxiang Zhao[1], Ping Lu[1,*]

[1]National Engineering Research Center of Next Generation Internet Access System, School of Optical and Electronic Information, Huazhong University of Science and Technology, Wuhan 430074, China

## Abstract

This paper designs and develops a miniaturized photoacoustic spectroscopy gas probe with acoustic pressure enhancement for in-situ detection in oil-immersed power equipment. The probe adopts a single-cavity single-fiber structure, integrating the photoacoustic cell and the Fabry-Pérot optical sensing cavity within a ceramic ferrule with an outer diameter of only 800 μm. Through the electro-mechanical-acoustic equivalent model analysis of the semi-open photoacoustic cell, the enhancement mechanism is revealed that reducing the radius of the photoacoustic cell can form a flat region in the acoustic pressure response. The CrAgAu composite metal diaphragm is fabricated by electron-beam evaporation, and the gas vents are precisely machined by focused-ion-beam etching, successfully realizing the preparation of the sensor prototype. Experimental results demonstrate that the sensor exhibits a flat acoustic response in the frequency range of 2800–10000 Hz with a sensitivity of –15 dB re 1 mV/Pa. Using acetylene as the target gas, the detection limit reaches 71.4 ppb (with an integration time of 130 s) at the optimal operating frequency of 3275 Hz, and the linearity exceeds 0.998. In-oil tests verify its stable detection capability in oil-phase environments, providing a miniaturized, high-sensitivity all-optical technical solution for in-situ monitoring of dissolved gases in transformer oil.

## 1. Introduction

Power transformers and other oil-immersed equipment are core to grid stability. Long-term faults generate dissolved gases ($H_2$, $C_2H_2$, $CH_4$, $C_2H_4$, CO) in oil, whose concentrations indicate insulation condition and fault severity, making DGA a key diagnostic tool [1-3]. GC is the gold standard for DGA[4-5] but is bulky, slow, and consumable-intensive, hindering real-time monitoring. TDLAS offers fast, selective, no-consumable detection [6-8], yet its ppm-level sensitivity falls short of ppb-level needs for early fault warning.

Photoacoustic spectroscopy (PAS) technology, as an indirect absorption spectroscopy method, retrieves gas concentration by detecting the acoustic signal generated after gas molecules absorb modulated light energy. Theoretically, it is a zero-background detection technology with significant advantages including high sensitivity, good selectivity, and fast response speed[9,10]. In recent years, with the rapid development of optical fiber sensing technology and micro-electro-mechanical system (MEMS) fabrication processes, photoacoustic spectroscopy gas sensors are continuously evolving toward miniaturization, all-optical integration, and integration. In 2022, the Chen Ke-Gong Zhenfeng team first reported an all-optical photoacoustic gas sensor based on a single-cavity single-fiber structure, which integrates the photoacoustic cell and the Fabry-Pérot (FP) optical sensing cavity into one, significantly reducing the sensor volume to below the millimeter scale[11]. Subsequently, the Ma Jun team further integrated the gas cell and the optical microphone monolithically on the end face of a single optical fiber, reducing the sensor diameter to only 125 μm and achieving a detection limit as low as 9 ppb for acetylene[12]. In 2025, the Wei Jin team from The Hong Kong Polytechnic University fabricated an optomechanical resonant spring structure on the end face of a fiber ferrule using 3D micro-nano laser printing technology, achieving trace acetylene detection under a high-frequency resonant operating mode at 105 kHz[13]. The above studies indicate that the miniaturization and all-optical integration of photoacoustic spectroscopy technology have become an important development trend in the field of trace gas sensing.

However, existing miniaturized photoacoustic spectroscopy sensors are mostly designed for open-air or simulated environments. For in-situ DGA in transformer oil, three major challenges remain: oil-sealing durability, ultra-compact size, and response speed. To address these issues, this paper develops an acoustically enhanced miniaturized photoacoustic spectroscopy gas probe specifically for in-situ detection in oil-immersed equipmentThe sensor adopts a single-cavity single-fiber structural scheme, integrating the photoacoustic cell and the Fabry-Pérot optical sensing cavity within a single ceramic ferrule with a maximum outer diameter of only 800 μm, achieving highly miniaturized and all-optically integrated sensing. By establishing an electro-mechanical-acoustic equivalent model of the semi-open photoacoustic cell, the intrinsic relationship between the photoacoustic cell dimensions and the acoustic frequency response is revealed. It is elucidated that reducing the radius of the photoacoustic cell enables the gas plane-wave damping time to dominate in the low-frequency region, thereby forming a flat frequency region in the acoustic pressure response—an acoustic pressure enhancement mechanism—and the influence of the vent hole size on the diaphragm on the acoustic characteristics of the photoacoustic cell is clarified. The CrAgAu composite metal diaphragm is fabricated by electron-beam evaporation, and precise machining of the gas vents is achieved through diaphragm transfer and focused-ion-beam etching techniques, successfully preparing the sensor prototype. Experimental results show that the sensor exhibits a flat acoustic response characteristic in the frequency range of 2800–10000 Hz with an acoustic pressure sensitivity of –15 dB re 1 mV/Pa. Using acetylene as the target gas, at the optimal operating frequency of 3275 Hz, the system achieves a detection limit of 71.4 ppb for acetylene (with an integration time of 130 s), and the linearity exceeds 0.998. The sensor combines miniaturized dimensions, high sensitivity, all-optical passive characteristics, and engineering practical potential, providing a new technical solution for in-situ detection of dissolved gases in power transformer oil and other trace gas monitoring scenarios in confined spaces.

## 2. Structure and principle

### A. Sensor Design and Simulation

As shown in Figure 1, this work designs a single-cavity single-fiber photoacoustic spectroscopy gas probe that integrates the photoacoustic cell and the optical Fabry–Pérot cavity into one entity (hereinafter referred to as the "gas probe").The non-resonant cylindrical photoacoustic cell of the "gas probe" is jointly constituted by the end face of a single-mode fiber, the CrAgAu composite metal diaphragm, and the inner wall of the ceramic ferrule. Meanwhile, the fiber end face and the composite diaphragm form the FP sensing cavity, and both the probe light and the pump light are transmitted through this single-mode fiber. Four small holes are arranged around the acoustic-sensitive region of the composite metal diaphragm as gas vents, which can simultaneously satisfy the dual requirements of efficient gas equilibrium and maintenance of the photoacoustic cell pressure. After the target gas enters the FP-type photoacoustic microcavity through the vents, it absorbs the wavelength-modulated pump light and generates a mixed-frequency photoacoustic signal. This signal drives the CrAgAu composite metal diaphragm to undergo forced vibration, causing a change in the cavity length. This change is then perceived by the probe light, and finally, the second-harmonic signal is extracted at the demodulation end to retrieve the gas concentration.

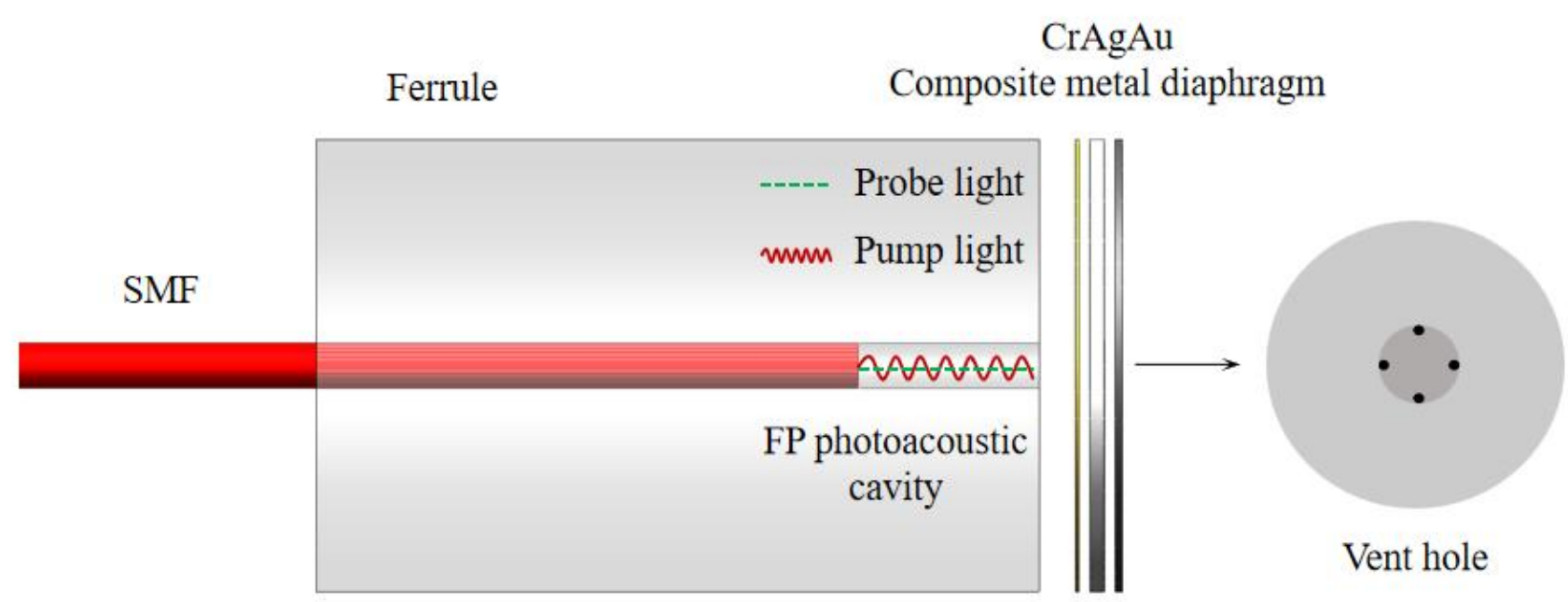

**Fig. 1.** Structural schematic of the single-cavity single-fiber all-optical gas probe.

The ceramic ferrule, which serves for protection and positioning, has an outer diameter of 800 μm and an inner diameter (i.e., the diameter of the FP-type photoacoustic microcavity) of 126 μm. The cavity length is designed to be 270 μm, but the actual cavity length depends on the optimal contrast point during the sensor fabrication process, generally with a deviation of no more than 20 μm. The acoustic-sensitive diaphragm is a 550-nm-thick CrAgAu (20nmCr + 500nmAg + 30nmAu) composite metal film. The centers of the vent holes on the diaphragm are distributed at the four quarter points of the photoacoustic cavity circumference, and the designed diameter of each vent hole is 6 μm. Since the sensor resembles a probe in shape, its miniaturized footprint (space occupation size) is measured by the maximum outer diameter of 800 μm.The total designed volume of the sensor does not exceed 0.5 $mm^3$, and the designed volume of the photoacoustic cell is approximately 0.003 $mm^3$.

The hard acoustic boundary formed by the ferrule inner wall tightly confines the photoacoustic wave, reducing the cavity radius, enhancing the pressure amplitude, and creating a flat frequency response region. In conventional low-frequency PAS systems, the degraded low-frequency response and high-frequency attenuation limit operation to low modulation frequencies. In contrast, the acoustically enhanced cell enables operation at higher frequencies, yielding improved SNR.

The constant $F_2$ of a non-resonant photoacoustic cell is generally defined as:

$$F_2 = \frac{\gamma - 1}{\omega \pi R^2} \frac{1}{\sqrt{1 + \left( \frac{1}{\omega \tau_0} \right)^2}} \tag{1}$$

The constant $F_1$ of a resonant photoacoustic cell decreases with increasing resonant frequency $\omega_j$. However, the resonant frequency of a small-volume FP-type photoacoustic cell often exceeds 10 kHz. If operated in the resonant mode, the amplitude of the photoacoustic signal would be significantly reduced. Therefore, the non-resonant operating mode is the dominant distribution pattern of the photoacoustic pressure in FP-type photoacoustic cells. For the constant $F_2$ of a non-resonant photoacoustic cell, Equation (1) can be transformed as:

$$F_2 = \frac{Q}{R^2} \frac{1}{\sqrt{\omega^2 + \frac{1}{\tau_0^2}}} \quad (2)$$

where $Q$ is a constant, and $R$ is the radius of the photoacoustic cell. When $\omega \ll 1/\tau_0$, Equation (2) can be expressed as:

$$F_2 = \frac{Q\tau_0}{R^2} \quad (3)$$

At this point, $F_2$ can be approximated as being independent of $\omega$. Therefore, reducing the radius of the photoacoustic cell enables a flat region to appear in the frequency response of the non-resonant photoacoustic pressure, allowing the sensor to operate at higher modulation frequencies and effectively overcoming the issue of significant low-frequency environmental noise. Furthermore, it can be seen from Equation (1) that the photoacoustic pressure amplitude in a non-resonant cell is inversely proportional to the cell radius. Hence, reducing the cell radius also directly enhances the photoacoustic pressure within the cavity.

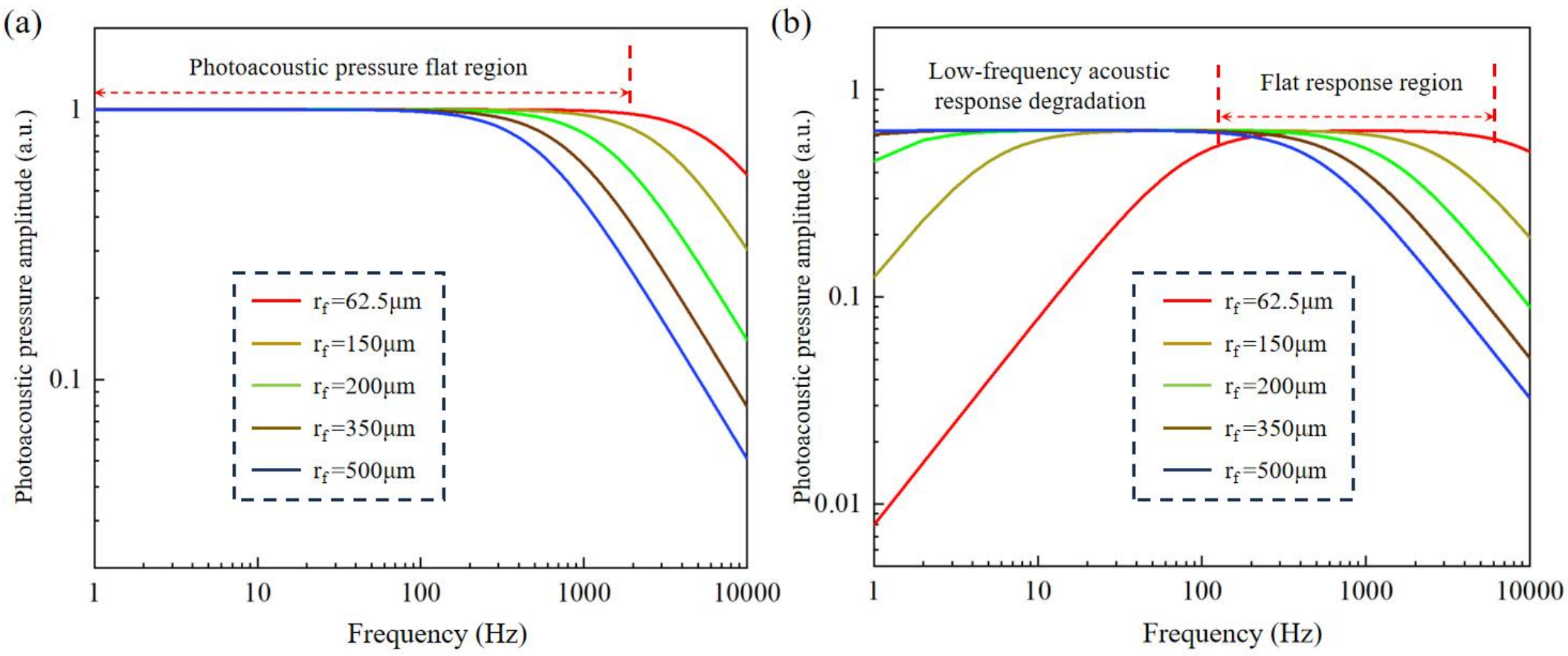

**Figure.2.** (a) Frequency characteristics of non-resonant photoacoustic pressure in photoacoustic cells with different radii; (b) Frequency response characteristics of photoacoustic cells with different radii.

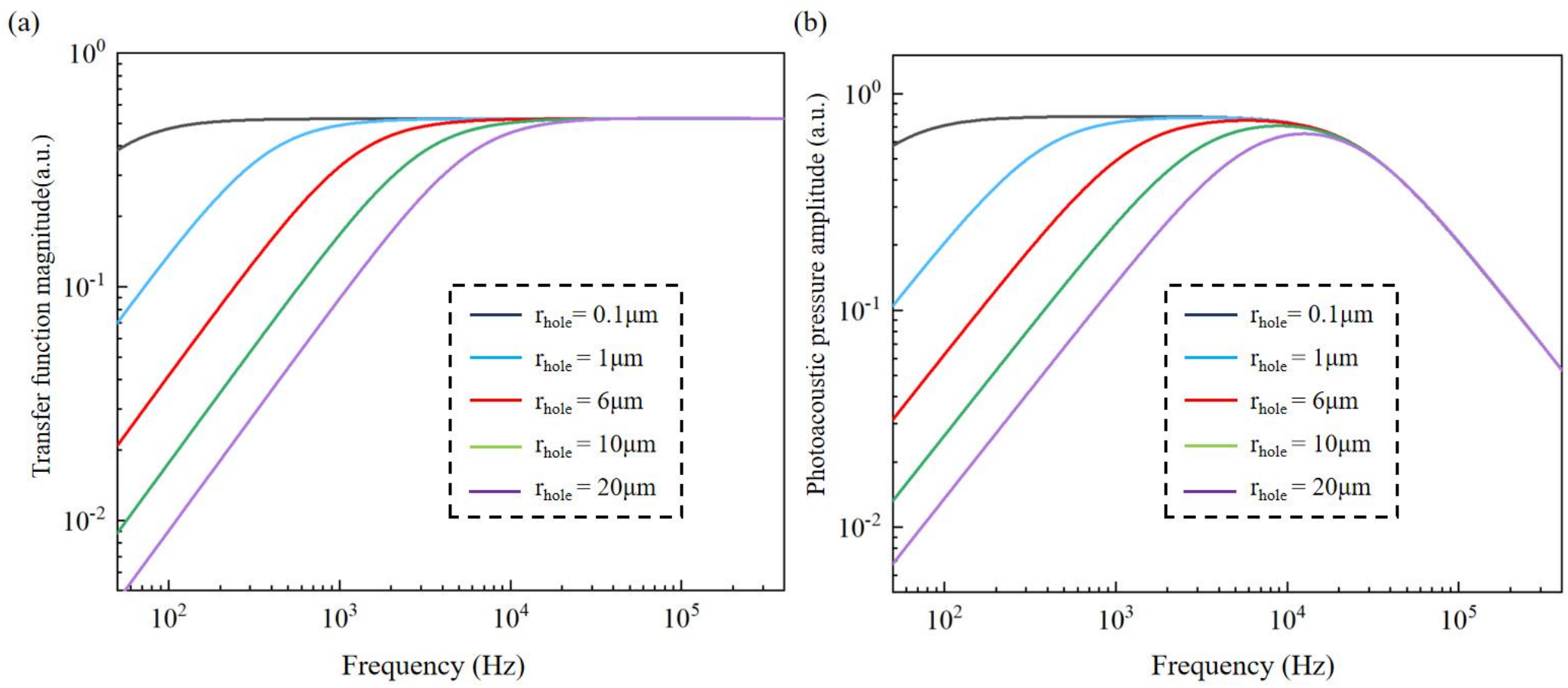

**Figure.3.** (a) Acoustic transfer functions under different vent-hole radii; (b) Photoacoustic response characteristics under different vent-hole radii.

Figure 2 simulates the frequency characteristics of the non-resonant photoacoustic pressure in photoacoustic cells with different radii. As can be seen from Figure 2(a), the smaller the radius of the photoacoustic cell, the shorter the plane-wave damping time, and the broader the flat frequency region, which effectively addresses the attenuation issue of the non-resonant photoacoustic pressure in the high-frequency region. To simulate the final frequency response characteristics of the photoacoustic pressure, the acoustic response characteristics of the photoacoustic cell must also be taken into account. Assuming an equivalent vent-hole radius of 6 μm and a length of 580 nm, the high-pass filtering effect introduced by the vent holes is incorporated, yielding the frequency response characteristics of photoacoustic cells with different radii, as shown in Figure 2(b). It can be observed that the flat response region of the photoacoustic cell is jointly determined by the non-resonant photoacoustic pressure characteristics and the high-pass filtering characteristics of the vent holes. Among these, the degradation of the low-frequency acoustic response determines the lower cutoff frequency of the flat response region, while the high-frequency attenuation characteristics determine the upper cutoff frequency of the flat response region.

Figure 3 presents the acoustic transfer function on the diaphragm compliance for different vent-hole radii, which is used to analyze the influence of the vent-hole size on the photoacoustic pressure amplitude inside the cavity, assuming a photoacoustic cavity radius of 62.5 μm and a vent-hole length of 580 nm. As shown in Figure 3(a), for the photoacoustic wave inside the cavity, the vent hole acts as a high-pass filter. A larger vent-hole radius results in a higher low-frequency cutoff frequency, thereby degrading the low-frequency response of the photoacoustic cavity. However, a larger vent hole also enables a faster gas exchange rate, shortening the response time of the sensing unit. In the design of the acoustically enhanced photoacoustic cavity, the optimal operating frequency has already been shifted to a higher region; therefore, the frequency characteristics of the photoacoustic pressure must also be taken into account. As shown in Figure 3(b), the vent-hole radius determines the lower cutoff frequency of the flat response region of the photoacoustic cavity. For a given cavity radius, an excessively large vent-hole radius can cause this flat region to disappear and even deteriorate the optimal signal-to-noise ratio of the photoacoustic cavity. Taking a vent-hole radius of 20 μm as an example, this dimension is already close to two-thirds of the cavity radius. Due to the low-pass filtering effect introduced by such a large opening, a significant portion of the non-resonant photoacoustic pressure in the 0–10000 Hz frequency range leaks out. At frequencies above 10000 Hz, which already exceed the flat region determined by the plane-wave damping time of the gas inside the cavity, the photoacoustic pressure drops sharply with increasing frequency, making it difficult to identify a suitable operating frequency. Therefore, in the design, a sufficient flat frequency region must be reserved for the photoacoustic pressure inside the cavity. In this work, a vent-hole radius of $r = 6$ μm is adopted. It should be noted that the vent-hole radii used in the simulation are all equivalent radii. For example, if four vent holes each with a radius of 3 μm are designed, the equivalent radius is 6 μm, corresponding to an equivalent total opening area.

## B. Sensor Fabrication

The rear end of the fiber ferrule is designed with a tapered through-hole to facilitate the free insertion of the single-mode fiber during device fabrication and cavity length adjustment. This tapered hole exists only in the rear region and does not affect the actual radius of the front photoacoustic cavity. The other end of the ferrule has a flat end face to facilitate the transfer of the composite metal diaphragm. The ferrule has an outer diameter of 0.8 mm and an inner diameter of 125 μm. To prevent frequent tipping during the diaphragm transfer process, the ferrule length is controlled to approximately 3 mm. The vent holes are arranged around the acoustic-sensitive region of the composite diaphragm to avoid affecting the reflection of the probe light from the central area of the diaphragm. Four short through-holes, each with a radius of 3 μm and extending through the entire diaphragm thickness, are fabricated. The fabrication of the acoustically enhanced photoacoustic probe mainly consists of two processes: the transfer of the composite metal diaphragm and the adjustment of the cavity length, as shown in Figure 4.

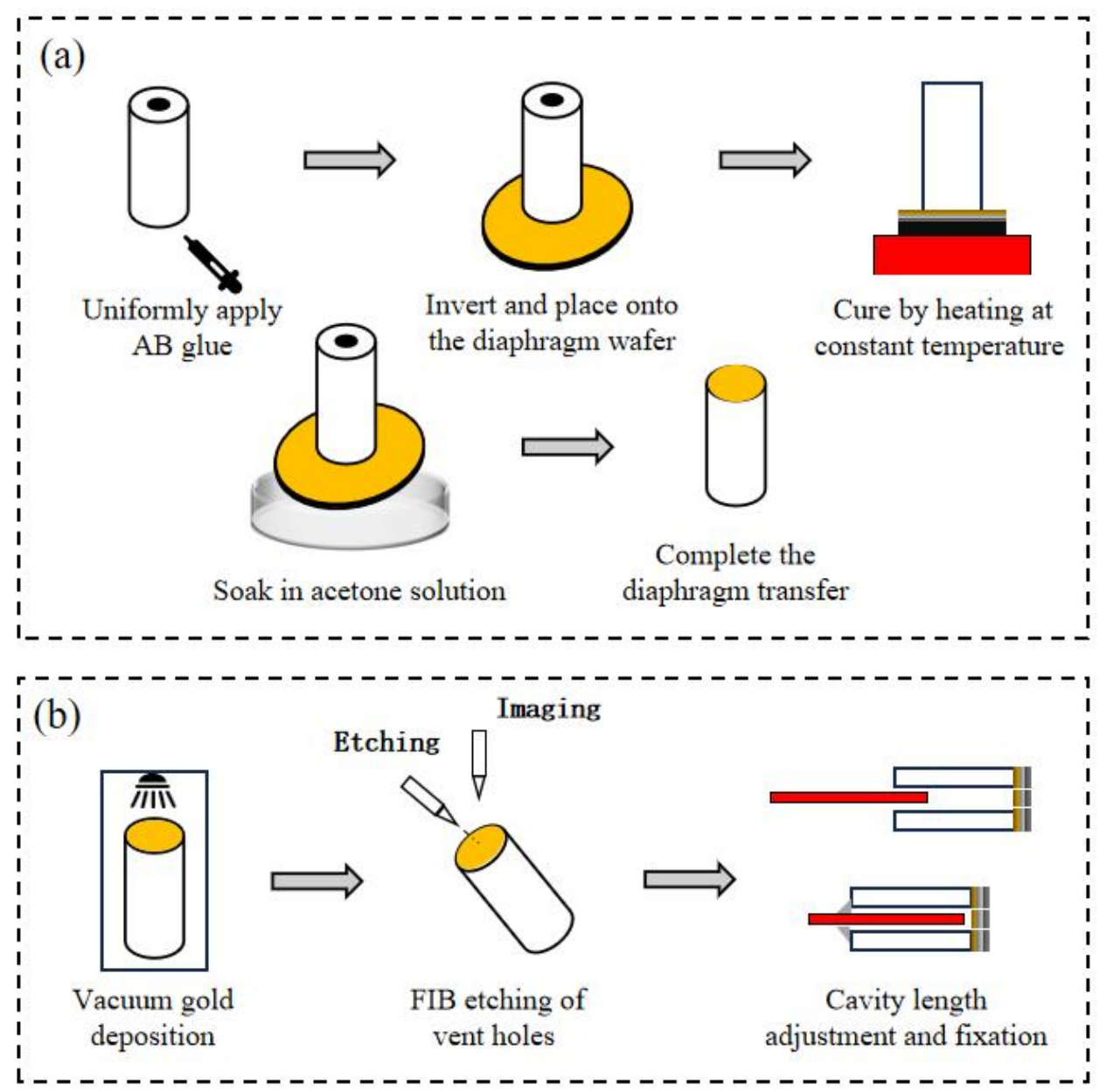

**Figure.4.** Fabrication process of the acoustically enhanced MPAS gas probe: (a) Transfer of the CrAgAu composite metal diaphragm; (b) Vent-hole etching and cavity length adjustment of the sensing unit.

The composite metal diaphragm was fabricated and transferred by sequentially depositing 30 nm Cr, 500 nm Ag, and 20 nm Au on a silicon substrate using electron-beam evaporation to form the Cr–Ag–Au composite film. The Cr layer serves as a sacrificial layer to facilitate subsequent lift-off, the Ag layer provides high reflectivity to ensure the interference contrast of the FP cavity, and the Au layer acts as an anti-oxidation protective layer. During the diaphragm transfer, 353ND epoxy resin AB adhesive was mixed at a ratio of 1:9. A small amount of the mixed adhesive was picked up with a single-mode fiber and uniformly spotted around the central hole on the flat end face of the ferrule. The fiber was then inserted through the through-hole to remove residual adhesive, preventing it from seeping into and affecting the photoacoustic cavity structure. Subsequently, the ferrule was inverted and placed onto the diced Cr–Ag–Au diaphragm wafer, and heated at 80 °C for 60–80 minutes to fully cure the AB adhesive. Finally, the cured ferrule and wafer were soaked in acetone for 24 hours to dissolve the bonding layer between the Cr layer and the silicon substrate, achieving complete transfer of the composite diaphragm onto the end face of the ferrule. The vent holes were fabricated by focused ion beam (FIB) etching. A gold film with a thickness of less than 5 nm was first deposited on the ferrule end face to enhance electrical conductivity and imaging accuracy. The sample was then placed into an electron-beam/focused-ion-beam dual-beam system, and the acoustic-sensitive region of the diaphragm (with a diameter of 125 μm) was precisely located by EBL imaging. Four vent holes with a radius of 3 μm were etched using an ion beam current of 15 nA. Finally, the single-mode fiber was inserted into the ferrule through-hole, and its position was precisely controlled using a six-axis alignment stage. The interference spectrum of the FP cavity was monitored in real time until the contrast reached 21.7 dB (corresponding to a cavity length of approximately 200 μm), after which the fiber was fixed by adhesive dispensing. The sensor fabrication was completed after 24 hours of stress relaxation. The single-mode fiber was precisely aligned with the ferrule using a six-axis alignment stage, positioned at the center of the ferrule and perpendicular to the rear end face before being inserted into the through-hole. The interference spectrum of the FP cavity was monitored in real time until the contrast reached 21.7 dB (cavity length ≈ 200 μm). The fiber was then fixed by adhesive dispensing and left to stand for 24 hours to release stress, completing the probe fabrication.

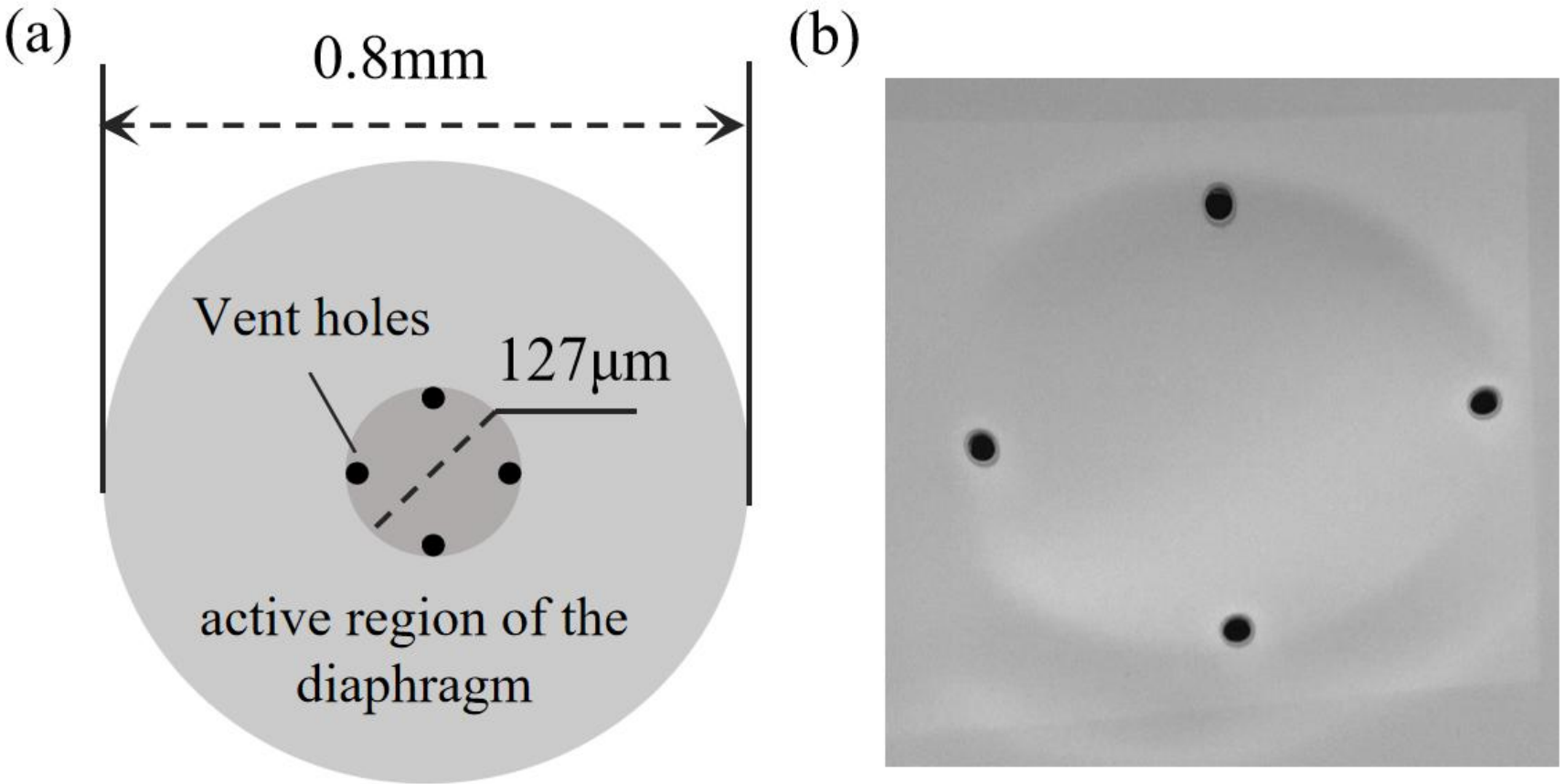

**Figure.5.** (a) Schematic diagram of the vent-hole dimensions; (b) SEM image of the composite diaphragm end face after etching.

## 3. Sensor Performance

### A. Test System

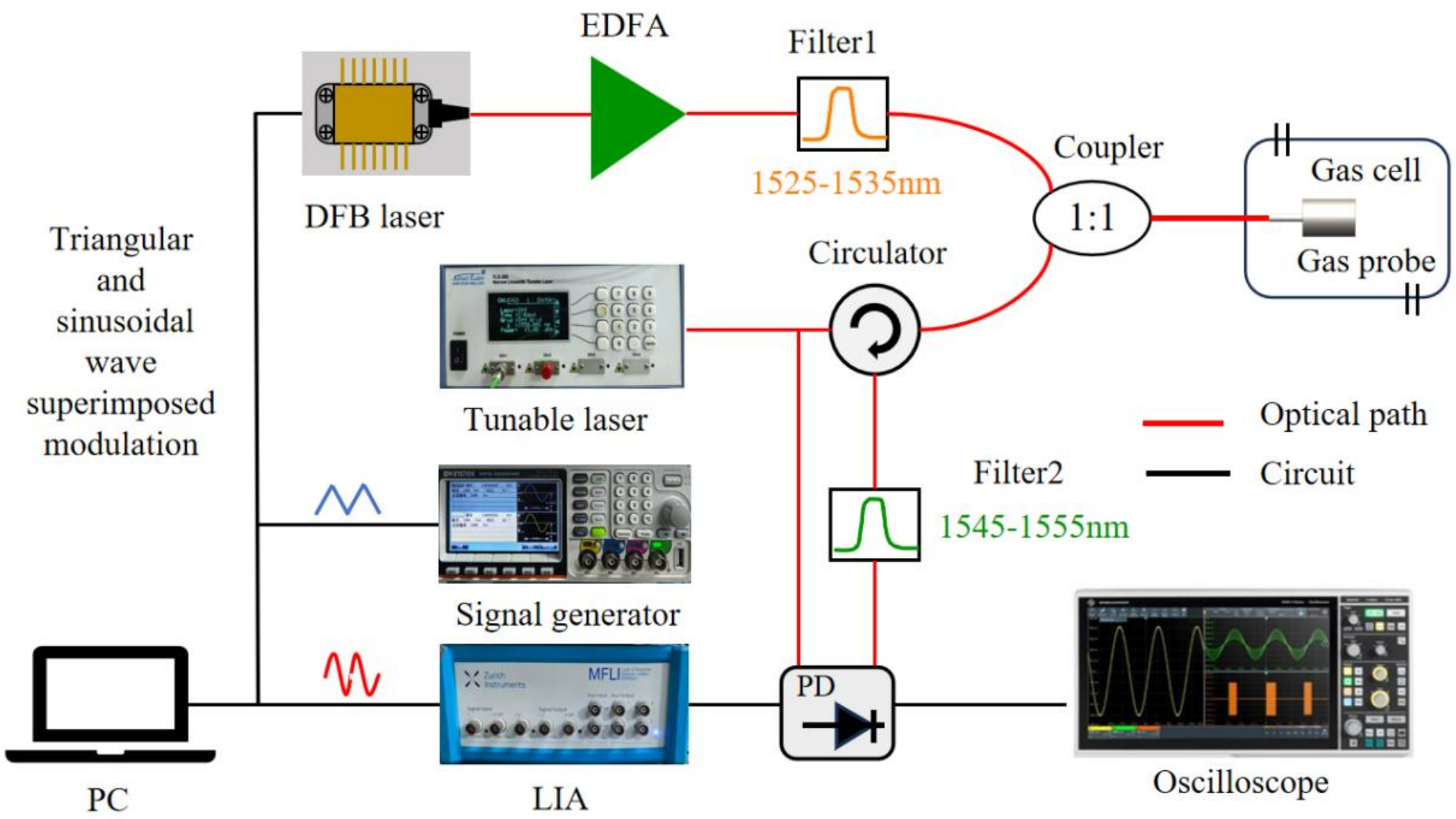

**Figure.6.** Gas sensing test system for the "gas probe".

To test the gas probe, a system was built (Fig. 6). A DFB laser (pump) was EDFA-amplified, combined with a tunable probe laser, and sent to the sensor. The return light was filtered, differentially detected, and split to an oscilloscope (for Q-point monitoring) and a lock-in amplifier (for 2f signal extraction). Modulation used a triangular wave (from signal generator) plus a sine wave (from lock-in), superimposed on the laser driver. $C_2H_2$ at 1531.6 nm was targeted. EDFA sidelobes above 1535 nm were blocked by a 1525–1535 nm filter to avoid crosstalk. Q-point demodulation was used, with real-time oscilloscope monitoring; flow rate was kept ≤50 ccm to protect the diaphragm. With 1000 ppm $C_2H_2$, DFB output was 70 mW, EDFA output 200 mW (110 mW at sensor input). The triangular wave was ±100 mV at 20 mHz; its amplitude set the scan range, optimized experimentally for 2.2×FWHM modulation depth. The lock-in integration time was 1 s; sine amplitude was set to 250 mV to avoid nonlinearity.

## B. Gas-Phase Test Results

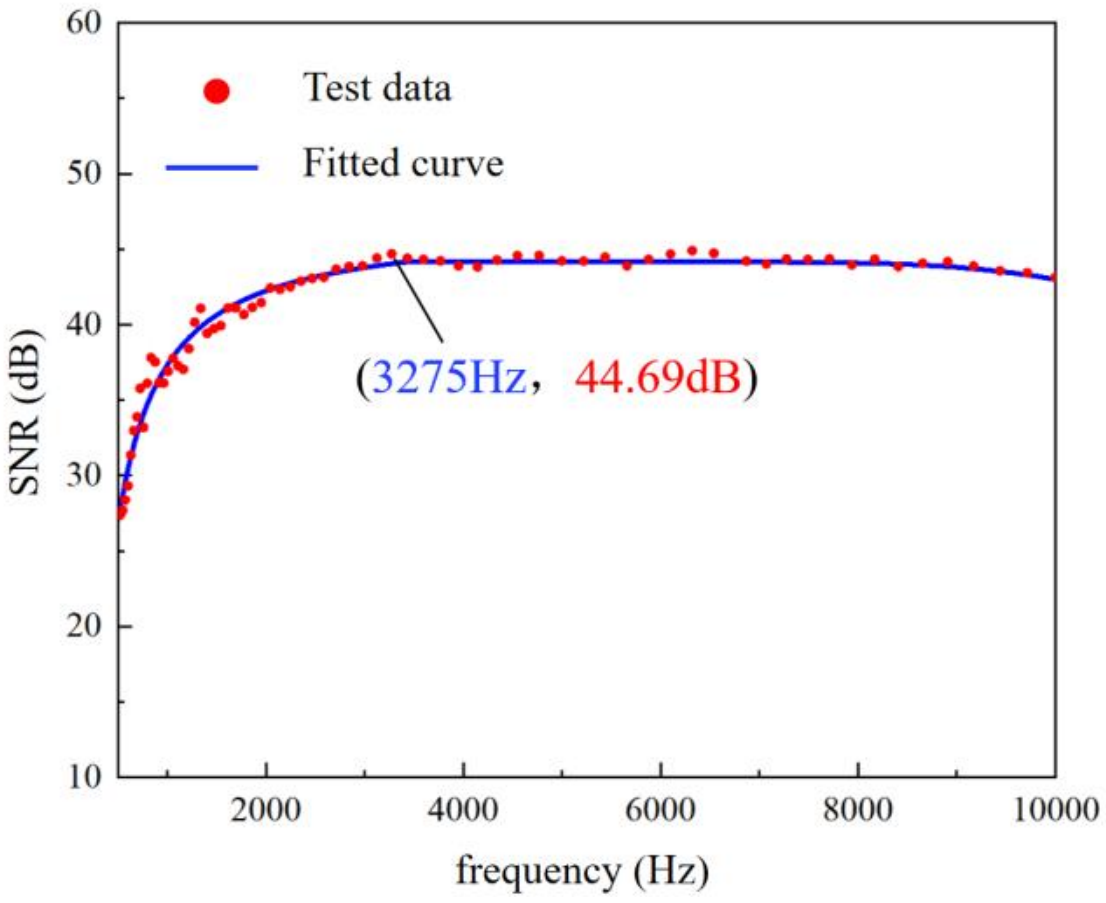

**Figure.7.** SNR frequency characteristic curve of the sensor from 800 to 10,000 Hz

The frequency range was set from 500 to 10,000 Hz. In the low-frequency region, the photoacoustic pressure response increased with increasing frequency, while in the high-frequency region, the signal amplitude exhibited a slow but continuous downward trend with increasing frequency. After the reduction of the photoacoustic cavity radius, the plane-wave damping time became dominant, and the acoustic pressure amplitude was approximately independent of frequency within the flat region. Experimentally, the sensor entered the flat region near 3125 Hz, and the highest signal-to-noise ratio (44.69 dB) was achieved at 3275 Hz.

The noise characteristics of the acoustically enhanced photoacoustic cell in the frequency range of 500–10,000 Hz are shown in Figure 7. The background noise in a pure nitrogen environment decreases slowly with increasing frequency. The resulting signal-to-noise ratio (SNR) frequency characteristics of the acoustically enhanced photoacoustic cell are also presented in the figure.

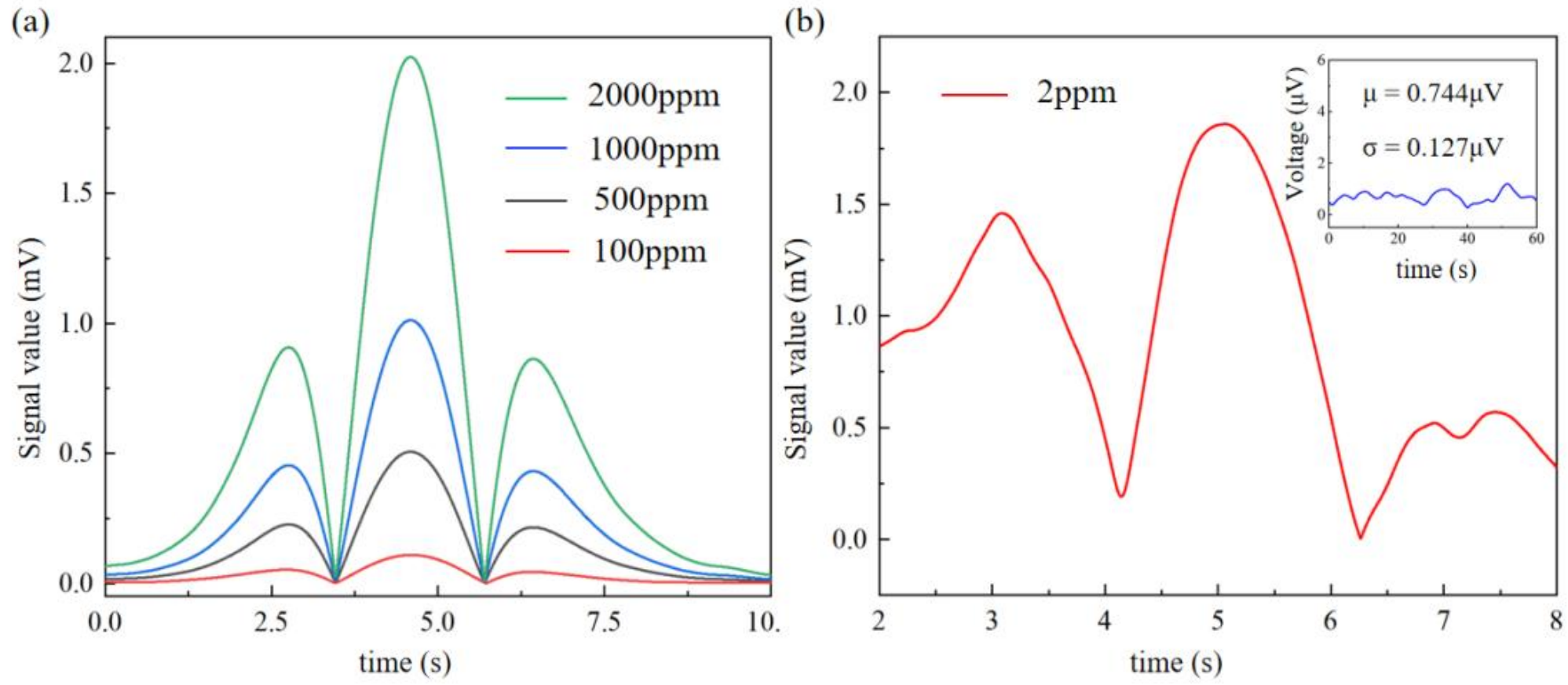

**Figure.8.** (a) Second-harmonic signals at different acetylene concentrations; (b) Second-harmonic signal for 2 ppm acetylene.

Figure 8(a) presents the second-harmonic waveforms of the acoustically enhanced photoacoustic probe when measuring acetylene at different concentrations, and Figure 8(b) shows the test results for 2 ppm acetylene. At a pump light modulation frequency of 3275 Hz, the noise mean value under pure nitrogen background was 0.744 μV, with a standard deviation of 0.127 μV.

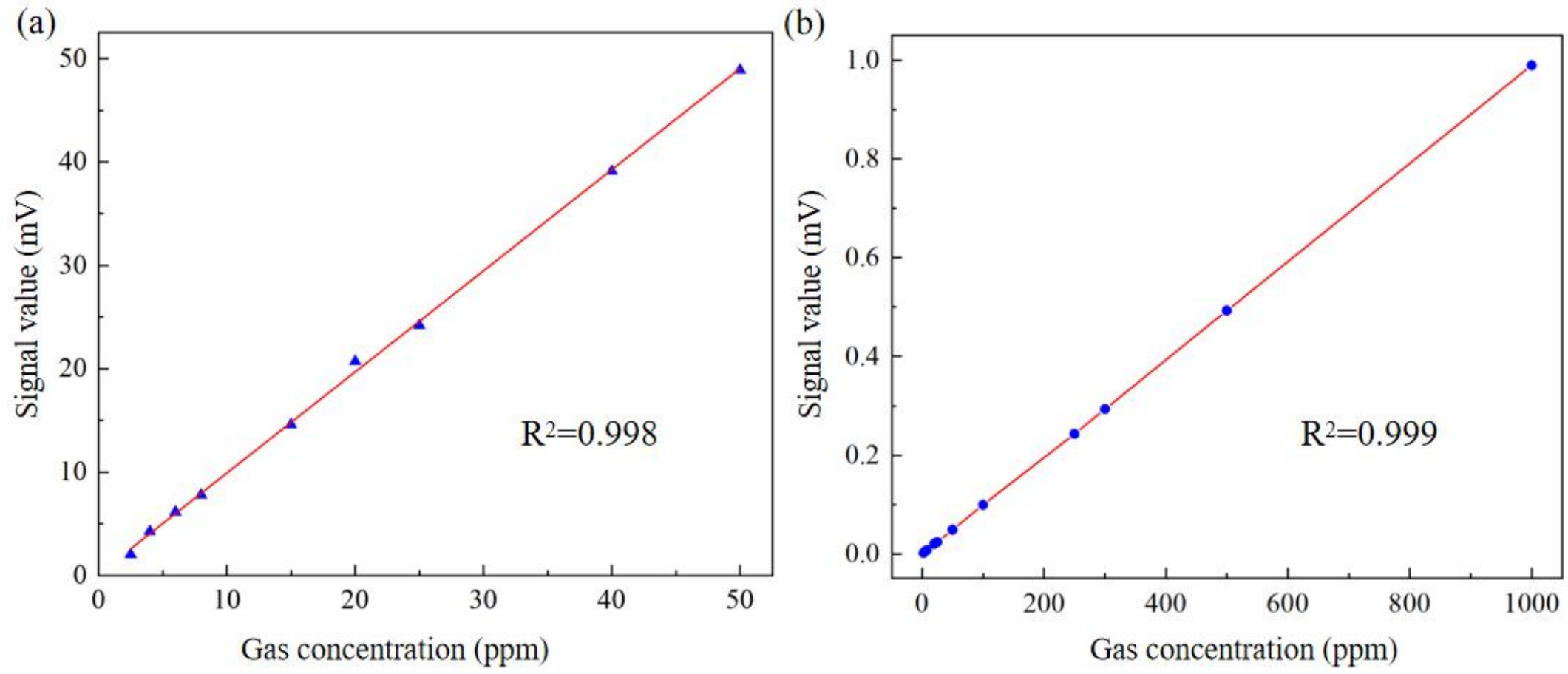

**Figure.9.** (a) Linearity test results in the range of 2–50 ppm; (b) Linearity test results in the range of 2–1000 ppm.

Figure 9 presents the linearity test results of the sensor in the concentration ranges of 2–50 ppm and 2–1000 ppm, with linearities ($R^2$) of 0.998 and 0.999, respectively. The response sensitivity obtained from the 2–50 ppm linearity test was 0.98 μV/ppm. From the noise standard deviation, the noise equivalent concentration (NEC) for acetylene was calculated to be 129.59 ppb.

To further investigate the noise level and stability of the acoustically enhanced photoacoustic probe, a 30-minute system noise test was conducted under a pure nitrogen background, followed by a detection limit analysis based on Allan variance, as shown in Figure 4-15. The detection limit at an integration time of 1 s was consistent with the NEC calculation results. The optimal integration time of the system was 130 s, at which the minimum Allan variance of the measurement noise was 0.0704 μV, yielding a minimum detection limit of 71.4 ppb @ 130 s. For integration times exceeding 150 s, the Allan variance exhibited a continuous upward trend, which is attributed to deterministic fluctuations in noise caused by drift of the sensor's static operating point, rendering the Allan variance-based method for suppressing high-frequency white noise ineffective.

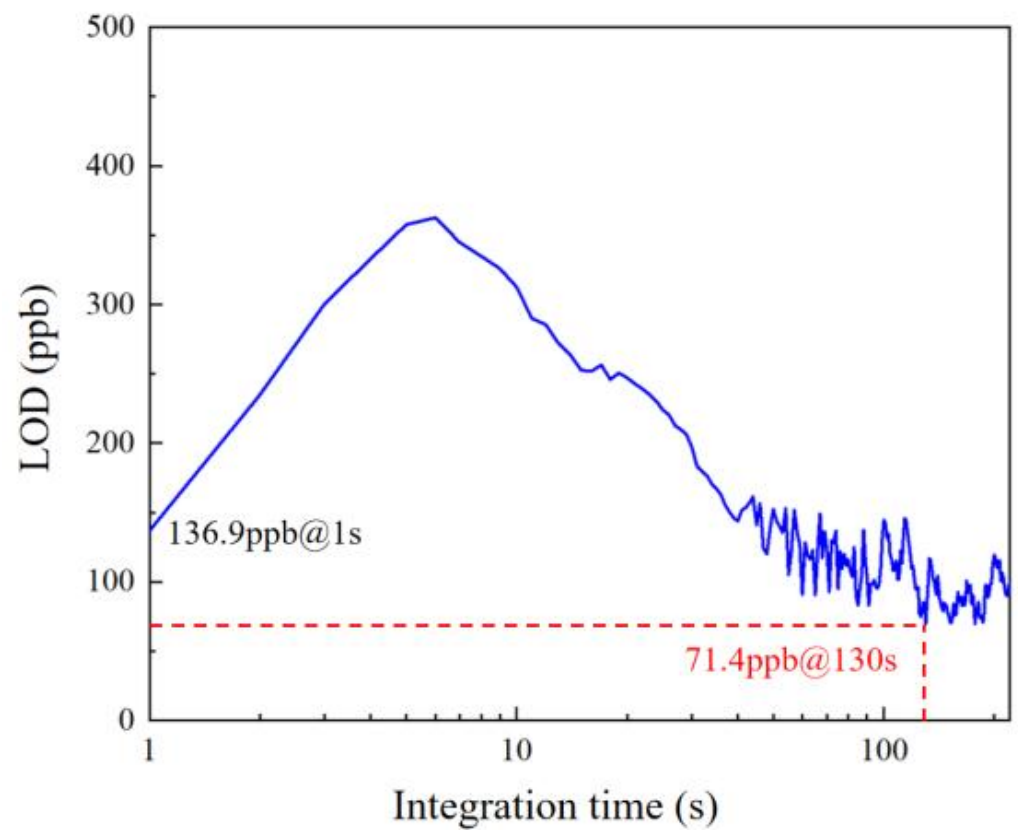

**Figure.10.** Detection limit analysis based on Allan variance under 1800 s noise test.

In the experiment, the incident pump optical power was 110 mW. The normalized noise equivalent absorption (NNEA) coefficient of the system was calculated to further evaluate the system sensitivity. The NNEA coefficient is defined as:

$$NNEA = \frac{\alpha P}{\sqrt{B}} \tag{4}$$

where α is the absorption coefficient of the gas at the system's minimum detection limit, P is the input optical power of the pump source, and B is the system bandwidth. The system bandwidth of the lock-in amplifier is 0.5Hz, and $P=110\ mW$. At an integration time of 130s, the NNEA coefficient of the "gas probe" for $C_2H_2$ is calculated to be $4.65\times10^{-10}$ W $cm^{-1}$ $Hz^{-1/2}$.

## C. In-situ Test of Acetylene in Insulating Oil

The oil – gas separation membrane can be placed in direct contact with the oil, enabling in-situ oil-gas separation via osmotic pressure, thus replacing conventional oil – gas separation units [13]. In this study, an AF2400 gas-permeable tube made of expanded polytetrafluoroethylene (ePTFE) was used to encapsulate the acoustically enhanced photoacoustic probe, as shown in Figure 11(a).

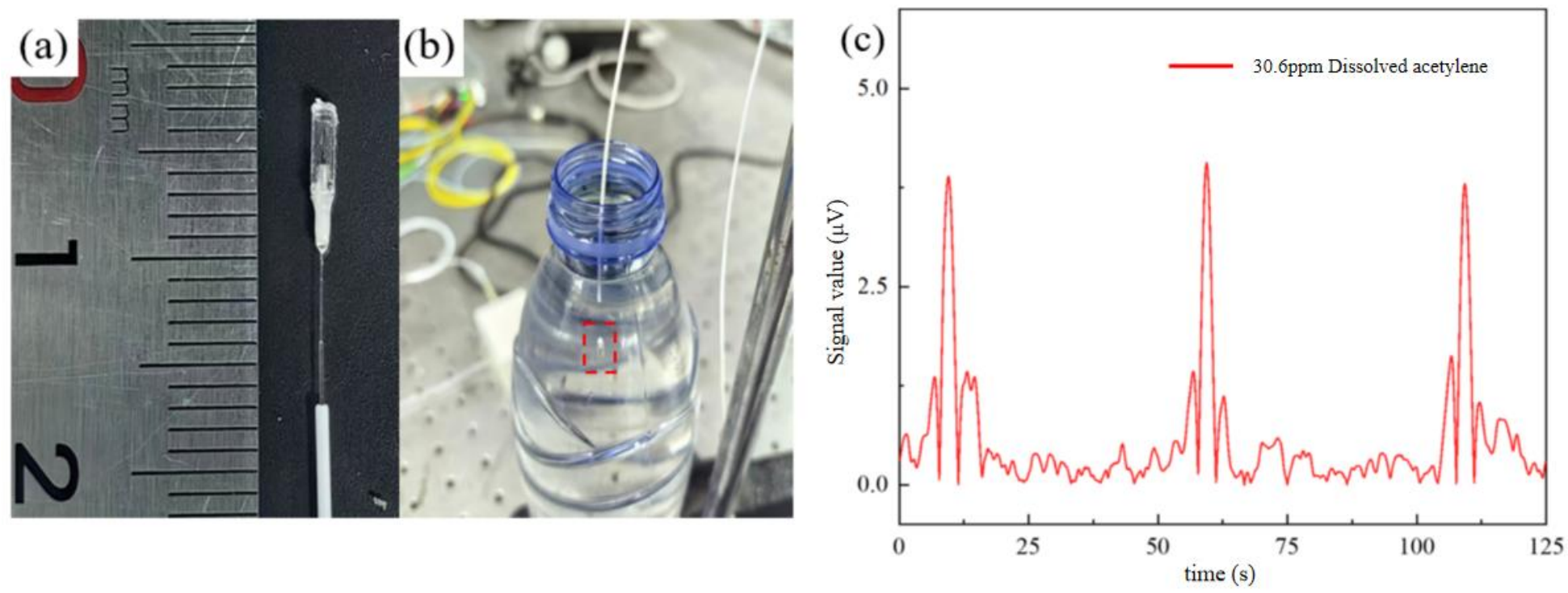

**Figure.11.** (a) MPAS gas probe encapsulated with the oil–gas separation tube; (b) Test of dissolved acetylene in oil; (c) Test results in 30.6 ppm oil sample.

Prior to the test, an oil sample with dissolved acetylene was prepared by bubbling 40 ppm standard acetylene gas into blank oil for two hours, during which magnetic stirring was continuously applied. After two hours, the oil sample was extracted and sealed for storage, and its concentration was calibrated as 30.6 ppm using an oil-phase gas chromatograph. The test results of the encapsulated photoacoustic probe in the oil-phase environment are shown in Figure 11(c). During the test, the input pump power was 35 mW, the modulation frequency was 3275 Hz, and the integration time was 1 s.

To observe the continuous testing capability and noise level of the sensor in the oil-phase environment, additional oil samples with 89.26 ppm and blank oil were prepared for testing. Figure 12 shows the variation curve of the second-harmonic signal when the sensor was moved from the 89.26 ppm oil sample to the 30.6 ppm oil sample over a 30-minute period. Subsequently, the sensor was placed in the gas-phase environment for 5 minutes and then re-immersed into blank oil for noise testing. The measured noise mean value was 0.795 μV, which is essentially consistent with the noise level under the nitrogen background.

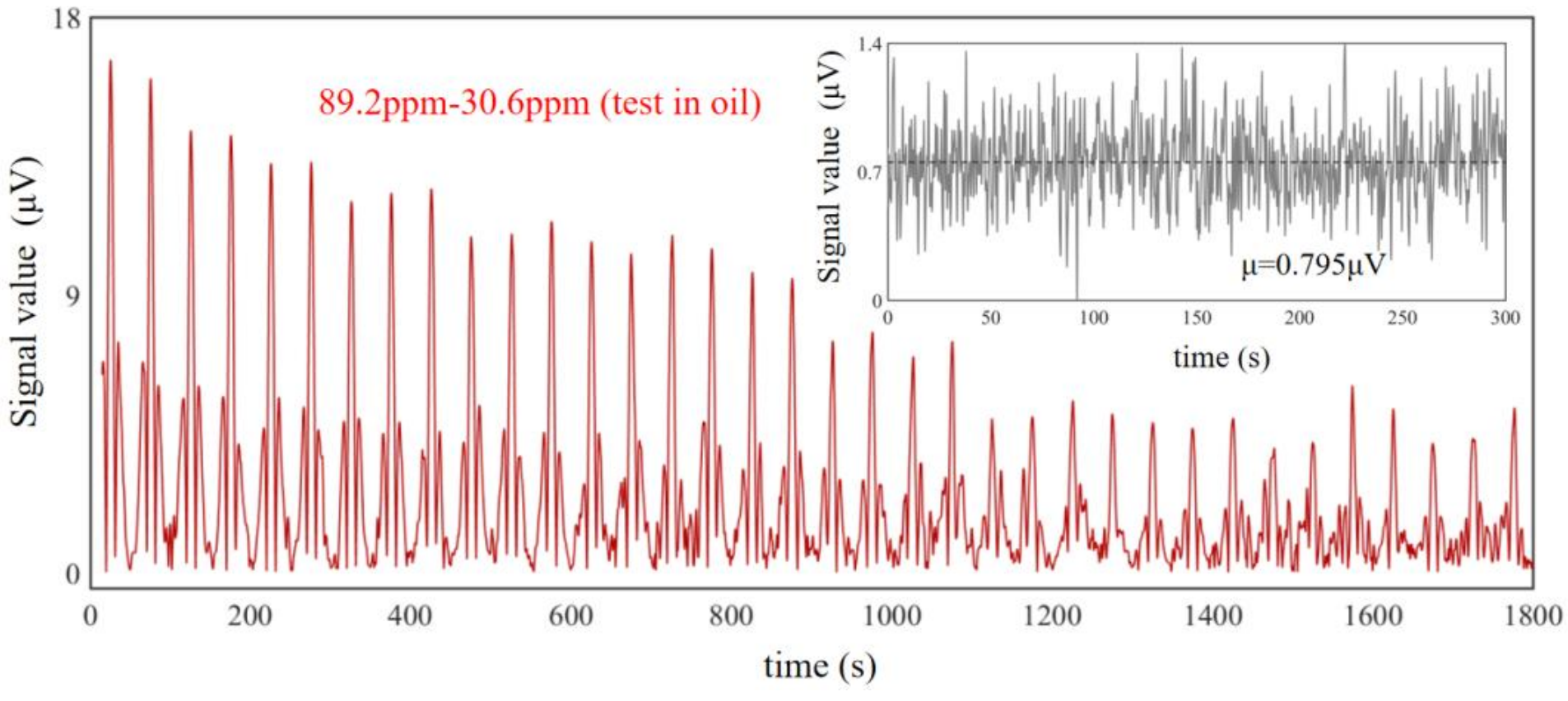

**Figure.12.** Continuous 30-minute test in the oil-phase environment from 90 ppm to 30 ppm.

## 4. Conclusion

In summary, this paper proposes a miniaturized photoacoustic spectroscopy gas probe with acoustic pressure enhancement, realized through a single-cavity single-fiber integration scheme, capable of ppb-level in-situ detection of dissolved acetylene in oil. Unlike conventional discrete photoacoustic cells or gas chromatography detection schemes, this design integrates the photoacoustic cell and the Fabry–Pérot optical sensing cavity into a single ceramic ferrule, which not only eliminates optical alignment errors but also reduces the maximum outer diameter of the sensor to 800 μm, achieving both miniaturization and all-optical integration. To address the inherent trade-off between acoustic pressure amplitude and operating frequency in non-resonant miniature photoacoustic cavities, this paper employs an electro-mechanical-acoustic equivalent model analysis, revealing that reducing the photoacoustic cavity radius (62.5 μm) allows the plane-wave damping time to dominate in the low-frequency region, thereby forming a flat acoustic pressure response region in the range of 2800–10000 Hz—an enhancement mechanism that effectively suppresses low-frequency environmental noise interference. Furthermore, by optimizing the vent-hole dimensions (radius 3 μm, four in total), the design maintains both efficient gas exchange rate and effective acoustic pressure accumulation within the photoacoustic cell. Experimental results demonstrate that the fabricated probe, with an outer diameter of 800 μm, achieves a detection limit of 71.4 ppb for acetylene at the optimal operating frequency of 3275 Hz (integration time 130 s), with a linearity better than 0.998, a response sensitivity of 0.98 μV/ppm across three concentration ranges, and a normalized noise equivalent absorption coefficient of $1.43 \times 10^{-8}$ $W \cdot cm^{-1} \cdot Hz^{-1/2}$. To the best of our knowledge, this is the smallest all-optical photoacoustic spectroscopy gas probe to date that possesses a flat broadband acoustic response and ppb-level sensitivity, and it is also the first miniaturized photoacoustic spectroscopy sensor that has been validated for in-situ operation in an oil-phase environment.

Given that research on miniaturized photoacoustic spectroscopy sensors capable of direct immersion in transformer oil remains extremely limited, this study provides key theoretical insights and engineering references for the design of next-generation in-situ monitoring probes for dissolved gases in oil. Moreover, its all-optical passive characteristics and sub-millimeter footprint facilitate integration with other sensors (e.g., temperature and pressure sensing units) in confined spaces, enabling the construction of multi-parameter fusion condition monitoring systems for transformers. The device exhibits a flat acoustic response over a wide frequency range of 2800–10000 Hz, and its noise level in the oil-phase environment (0.795 μV) is essentially consistent with that under gas-phase background. It offers significant advantages in the detection of weak photoacoustic signals and in-situ measurement in complex media, providing a new technical solution for applications such as early fault warning of power equipment, subsea transformer monitoring, and dissolved gas analysis in aerospace hydraulic systems.